\documentclass[12pt,preprint]{aastex} \usepackage{epsfig}
\usepackage{amsmath, amssymb,bm} \shorttitle{ Gravo-Magneto Limit
  Cycle} \shortauthors{Martin \& Lubow}
\begin{document}

\title{The Gravo-Magneto Limit Cycle in Accretion Disks}

\author{R. G. Martin \altaffilmark{}  \& S. H. Lubow \altaffilmark{} }
\affil{Space Telescope Science Institute, Baltimore, MD 21218 }
\date{}

\begin{abstract}
Previous theoretical studies have found that repeating outbursts can
occur in certain regions of an accretion disk, due to sudden
transitions in time from gravitationally produced turbulence to magnetically
produced turbulence.  We analyze the disk evolution in a state diagram
that plots the mass accretion rate versus disk surface density.  We
determine steady state accretion branches that involve gravitational
and magnetic sources of turbulence.  Using time-dependent numerical
disk simulations, we show that cases having outbursts track along a
nonsteady 'dead zone' branch and some steady state accretion
branches. The outburst is the result of a rapid inter-branch
transition.  The gravo-magneto outbursts are then explained on this
diagram as a limit cycle that is analogous to the well-known S-curve
that has been applied to dwarf nova outbursts.  The diagram and limit cycle provide a
conceptual framework for understanding the nature of the outbursts
that may occur in accretion disks of all scales, from circumplanetary
to protoplanetary to AGN accretion disks.
\end{abstract}

\keywords{accretion, accretion disks --- magnetohydrodynamics (MHD)
  --- planets and satellites: formation --- protoplanetary disks ---
  stars: pre-main sequence --- galaxies: nuclei}

\section{Introduction}

Disk turbulence plays a key role in the outward transport of angular
momentum that permits disk accretion.  Two main sources of disk
turbulence are gravitational instability \citep{paczynski78,lodato04} and
magnetic instability due to the magneto-rotational instability
\citep[MRI;][]{balbus91}.  For the magnetic instability to operate, a
critical level of ionization is required for the gas to couple
strongly enough to the magnetic field.  In certain situations, a
nonturbulent, high density, cool, weakly ionized disk layer known as a
'dead zone', can become gravitationally unstable, while remaining
magnetically stable \citep{gammie96}.  The gravitational instability
can lead to turbulent heating, resulting in increased ionization of
the gas. The ionization can then trigger the magnetic instability,
causing a much higher level of disk turbulence and accretion, an
outburst. After an outburst, the remaining disk gas cools and is
replenished by accreting gas from larger radii
\citep[e.g.][]{armitage01,zhu09}.  The outburst then repeats at later
times in an approximately periodic manner. Such behavior provides a
possible model for FU Ori outbursts in young stars.
  
The gravitational instability is thought to be a natural outcome of
dead zones in a layered disk, regions where the disk is nonturbulent
near the disk mid-plane, but turbulent due to the MRI near the disk
surface \citep{gammie96}.  The surface disk turbulence occurs as a
consequence of surface ionization by external sources of ionization,
such as cosmic rays or X-rays, that then permit the MRI to operate
\citep{glassgold04}.  In a layered region containing a dead zone,
steady state accretion is generally not possible and the dead zone
gains mass from some of the accretion flow near the disk surface.  As
mass builds up in the dead zone, it can become self-gravitating. As
discussed above, the self-gravitating state is turbulent and can
trigger a disk outburst.

The purpose of this {\it Letter} is to describe the disk outburst
cycle in terms of transitions between steady state configurations of
an accretion disk.  We determine the limit cycle for the outbursts in
a diagram that plots the mass accretion rate $\dot{M}$ versus disk
surface density $\Sigma$.  A similar approach was taken to explain the
disk instability in dwarf novae \citep{bath82, faulkner83}. The dwarf
nova thermal-viscous instability can be understood by the S-shaped
curve of steady state configurations in such a diagram.  In this
paper, we determine an analogous curve for outbursts in layered disks
with gravitational and magnetic sources of turbulence.  Our results do
not conform to an S-curve because the middle portion of the S is
missing.  That is, unlike the S-curve case, there is range of
accretion rates for which there are no steady state accretion
configurations.
 
 In Section~\ref{model} we describe the equations for the disk
 model. In Section~\ref{steady} we find steady state solutions to the
 disk equations and the resulting $\dot{M}$ versus $\Sigma$ curves. In
 Section~\ref{num} we describe the results of numerical simulations
 and show that the instability cycle can be understood in terms of
 transitions between the stable solutions in the $\Sigma$--$\dot{M}$
 diagram. Section~\ref{discussion} contains the discussion and
 conclusions.

\section{Disk Model}
\label{model}

The surface density evolution in an accretion disk is determined by
mass and angular momentum conservation \citep{pringle81}. The disk
model we adopt is essentially the same as Armitage et al. (2001).  
The main difference is that we
explicitly include the effects of both MRI and  self-gravity together
in the disk evolution equations, while they treated these effects
 in separate equations.  We have verified that the
numerical results that they obtained can be recovered using the
equations we describe below.

We consider a disk in Keplerian rotation with angular velocity
$\Omega=\sqrt{G M/R^3}$ around a mass $M$ at radius $R$. The disk
turbulence is modeled by the effective viscosities due to MRI and self-gravity.
The surface density evolution equation is
\begin{equation}
\frac{\partial \Sigma}{\partial t}=\frac{3}{R}\frac{\partial}{\partial
  R}\left\{ R^\frac{1}{2}\frac{\partial}{\partial R}\left[ 
\left( \nu_{\rm m} \Sigma_{\rm m} + \nu_{\rm g} \Sigma_{\rm g} \right) R^\frac{1}{2}\right]\right\},
\label{sigma}
\end{equation}
where $\nu_{\rm m}$ is the kinematic viscosity in the magnetic layers and $\nu_{\rm g}$
 is the kinematic viscosity 
due to turbulence associated with self-gravity of the gas that may act outside the MRI layers. 
 $\Sigma_{\rm m}$ is the surface density in the MRI active layers and
\begin{equation}
\Sigma_{\rm g}= \Sigma-\Sigma_{\rm m}
\label{sigma_g}
\end{equation} is 
the surface density 
outside the MRI active layers. $\Sigma_{\rm g}$ can contribute to the mass flux 
if  $\nu_{\rm g} > 0$.

We assume that the disk surface layers are ionized by external sources
to a maximum disk density depth of $\Sigma_{\rm crit}/2$ on the upper
and lower disk surfaces.  Therefore, the disk surface layers always
contain MRI turbulence.  At a given radius, we assume the disk is
sufficiently thermally ionized for MRI to operate throughout the
vertical extent of the disk if the mid-plane temperature is above some
critical value $T_{\rm crit}$.  Therefore, if either $T_{\rm c}>T_{\rm
  crit}$ or if $\Sigma<\Sigma_{\rm crit}$ at some radius, then the
disk is MRI turbulent (active) at all heights, where $T_{\rm c}$ is
the temperature at the disk mid-plane (central temperature).
Otherwise, there is either a dead zone layer or the disk is
self-gravitating in the presence of magnetic surface layers.

The temperature at the disk mid-plane $T_{\rm c}$ evolves according to the
simplified energy equation
\begin{equation}
\frac{\partial T_{\rm c}}{\partial t}=\frac{2(Q_+-Q_-)}{c_{\rm p}
  \Sigma}
\label{temp}
\end{equation}
\citep{pringle86,cannizzo93}.  The disk specific heat for temperatures
around $10^3\,\rm K$ is $c_{\rm p}=2.7 {\cal R}/\mu$, where ${\cal R}$
is the gas constant and $\mu$ is the gas mean molecular weight.  The
local heating owing to viscous dissipation is taken as
\begin{equation}
Q_+=\frac{9}{8}\Omega^2 \left( \nu_{\rm m} \Sigma_{\rm m} + \nu_{\rm
  g} \Sigma_{\rm g} \right).
 \label{Qp}
\end{equation}
To determine the local cooling rate, we assume that each annulus of the disk
radiates as a black body so that
\begin{equation}
Q_-=\sigma T_{\rm e}^4,
\end{equation}
where $T_{\rm e}$ is the temperature at the surface of the disk and
$\sigma$ is the Stefan-Boltzmann constant.

The kinematic turbulent viscosity in the magnetic layer is taken to be
\begin{equation}
\nu_{\rm m}=\alpha_{\rm m}\frac{c_{\rm m}^2}{\Omega},
\label{nua}
\end{equation}
where the sound speed is $c_{\rm m}=\sqrt{{\cal R} T_{\rm m}/\mu} $
with temperature in the magnetic layer $T_{\rm m}$.  The disk is
self-gravitating if the Toomre parameter $Q<Q_{\rm crit}$, where
\begin{equation}
Q=\frac{c_{\rm g}\Omega}{\pi G \Sigma},
\end{equation} 
and the sound speed at the disk mid-plane is given by $c_{\rm
  g}=\sqrt{ {\cal R}T_{\rm c}/\mu},$ where we approximate the temperature
of the self-gravitating region that extends to the disk mid-plane
as $\simeq T_{\rm c}$.  The effective kinematic viscosity from the
turbulence induced by the self-gravitational instability is
approximated by
\begin{equation}
\nu_{\rm g}=\alpha_{\rm g}\frac{c_{\rm g}^2}{\Omega}\left[\left( \frac{Q_{\rm crit}}{Q}\right)^2-1\right]
\label{nug}
\end{equation}
 for $Q<Q_{\rm crit}$ and zero otherwise
 \citep{lin87,lin90}.

The mid-plane disk temperature for an optically thick disk in thermal
equilibrium is obtained by considering the energy balance
in a  layered model above the disk mid-plane. One layer contains the surface density
$\Sigma_{\rm m}/2$. The other layer contains the complementary  surface density
$\Sigma_{\rm g}/2$. 
The results
are that
\begin{equation}
\sigma T_{\rm c}^4 = \frac{9}{8}\Omega^2 \left( \nu_{\rm m} \Sigma_{\rm m}\tau_{\rm m} + \nu_{\rm g} \Sigma_{\rm g} \tau \right)
\label{tau}
\end{equation}
and 
\begin{equation} 
T_{\rm m}^4=\tau_{\rm m}T_{\rm e}^4.
\label{rel}
\end{equation} 
The optical depth to the magnetic region is
\begin{equation}
\tau_{\rm m}=\frac{3}{8}\kappa(T_{\rm m}) \frac{\Sigma_{\rm m}}{2}
\end{equation}
and  the optical depth within the complementary region is
\begin{equation}
\tau_{\rm g}=\frac{3}{8}\kappa(T_{\rm c})\frac{\Sigma_{\rm g}}{2}
\end{equation}
with
\begin{equation}
\tau = \tau_{\rm m} + \tau_{\rm g}.
\end{equation} 
 Note $\tau_{\rm g}$ is defined even in a dead zone layer ($\nu_{\rm g} = 0$) with $\Sigma_{\rm g}$ defined by equation (\ref{sigma_g}).
We adopt the simplified opacity of Armitage et al. (2001)
\begin{equation}
\kappa(T)=0.02\,T^{0.8} cm^2/g.
\label{kappa}
\end{equation}
The energy equation~(\ref{temp}) in a steady state has the solution
\begin{equation}
\sigma T_{\rm e}^4= \frac{9}{8} \Omega^2 \left( \nu_{\rm m} \Sigma_{\rm m} + \nu_{\rm g} \Sigma_{\rm g} \right).
\label{heat}
\end{equation}
From equations (\ref{tau})-(\ref{heat}), we obtain an expression for the cooling function 
\begin{equation}
Q_-=\sigma T_{\rm e}^4 = \tau^{-1} \left( \sigma T_{\rm c}^4 + \frac{9}{8} \Omega^2 \nu_{\rm m}\Sigma_{\rm m}  \tau_{\rm g} \right).
\label{Qmg-}
\end{equation}
We apply this cooling
function to equation (\ref{temp}), even when the disk is not in
thermal equilibrium. This means that we do not attempt to treat the
cooling during the transitions consistently.

\section{Steady State Disks and State Transitions}
\label{steady}

We consider the disk to be supplied by material at a constant mass
input rate into an outer region. In a steady state, the mass flux
through the disk is constant in radius interior to the mass input
region.  We determine the relationship between the mass flux $\dot{M}$
and the disk surface density $\Sigma$ at some radius. 

In a steady state, far from radius of the central object,
the surface density equation~(\ref{sigma}) 
has the solution
\begin{equation}
\dot M=3 \pi \left( \nu_{\rm m} \Sigma_{\rm m} + \nu_{\rm g} \Sigma_{\rm g}\right).
\label{surf}
\end{equation}
From equations~(\ref{heat}) and~(\ref{surf}), we find the steady state
surface temperature of the disk
\begin{equation}
T_{\rm e}=\left(\frac{3\dot M \Omega^2}{8\pi\sigma }\right)^\frac{1}{4}.
\label{te}
\end{equation}
Notice that $T_{\rm e}$ is a function of the total accretion rate
$\dot{M}$.

\begin{figure}
\begin{center}
\includegraphics[width=8.8cm]{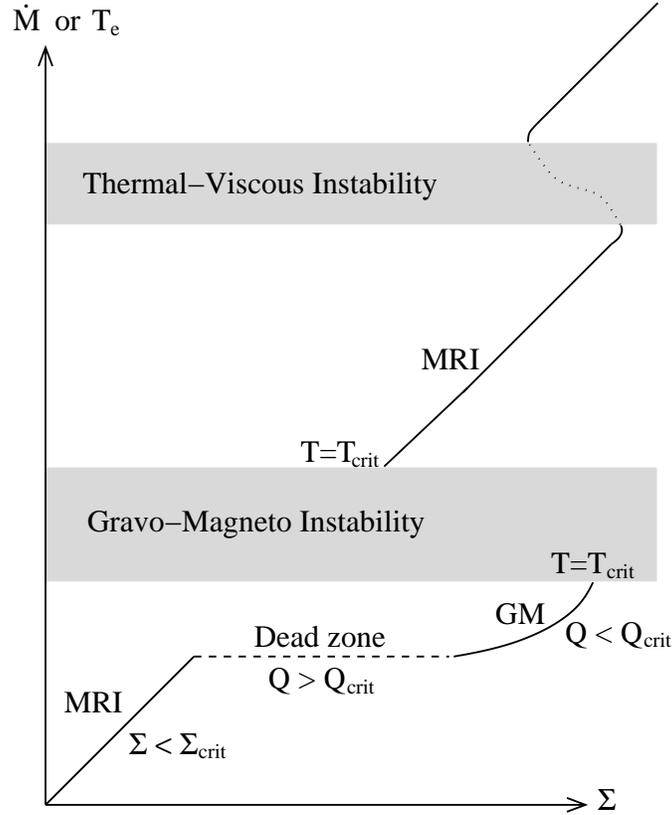}
\end{center}
\caption{A schematic diagram of steady disk solutions in the
  $\Sigma$--$\dot{M}$ plane at some radius in the disk. The solid
  lines show steady state solutions.  The two lower straight line
  branches on this log-log plot show the disk configurations with
  magnetic (MRI) turbulence at all heights.  The lower branch ends
  where $\Sigma=\Sigma_{\rm m}=\Sigma_{\rm crit}$. The higher branch
  begins where $T_{\rm c}=T_{\rm crit}$.  The curved GM branch is for
  a disk configuration with magnetically driven surface turbulence
  that overlies a mid-plane region that is turbulent due to the
  effects self-gravity.  The dashed line corresponds to nonsteady
  configurations in which a dead zone (nonturbulent region) underlies
  surface layers with MRI turbulence.  The lower shaded region
  contains no steady state solutions for any value of $\dot{M}$.  The
  upper S-curve describes the thermal-viscous instability. The dotted
  line shows its unstable steady state solutions. }
\label{diagram}
\end{figure}

We sketch typical solutions, in the $\Sigma$--$\dot{M}$ plane at a
given radius in Fig.~\ref{diagram} to illustrate their principal
properties.  We show the various branches of disk solutions, as
described below. In Section~\ref{num}, we consider particular
numerical solutions.

\subsection{MRI Disk Branches}

If $T_{\rm c}>T_{\rm crit}$ or $\Sigma<\Sigma_{\rm crit}$ at a given
radius, then the disk is assumed to be fully MRI turbulent, that is
magnetically turbulent at all heights ($\Sigma_{\rm g}=0$).
The disk surface density in this case simply follows from
equation (\ref{surf})
\begin{equation}
\dot M=3 \pi \nu_{\rm m}\Sigma_{\rm m}.
\label{1}
\end{equation}
 At a fixed radius, equations~(\ref{nua}),  (\ref{rel}), (\ref{te}), and 
(\ref{1})  imply that $\dot{M} \propto \Sigma_{\rm m}^{21/11}$.

\subsection{Gravo-Magneto(GM) Disk Branch}

Over a certain range of accretion rates, there exists a branch of
steady state solutions where the disk is gravitationally unstable and
has MRI active surface layers with $\Sigma_{\rm m}=\Sigma_{\rm crit}$.  We numerically determine
$\dot{M}(\Sigma)$ over a range of $\Sigma$ for which $Q < Q_{\rm
  crit}$ and $T < T_{\rm crit}$.  Equations (\ref{tau}) and
(\ref{surf}) are numerically solved together with auxiluary equations
(\ref{nua}) - (\ref{nug}), (\ref{rel}) - (\ref{kappa}), (\ref{heat}),
and (\ref{te}) to obtain this function.

 \subsection{Dead Zone Branch}
\label{dzone}

Between the lower MRI and the GM steady state branches there are disk
configurations that involve dead zones. The dead zone is a
nonturbulent mid-plane layer that lies vertically between MRI actively
accreting surface layers. Unlike the other branches we have
considered, this branch is not in steady state and therefore not
described by steady state solutions to the disk equations. It is
plotted as a dashed line in lower portion of Fig.~\ref{diagram}. If a
disk lies along this line, it will evolve to the right, as the dead
zone gains mass, while the disk accretion rate is pinned to the rate
provided by MRI turbulence that resides in surface layers of fixed
density $\Sigma_{\rm crit}$. The mass gain of the dead zone is
produced by some of the accretion flow in the disk surface layers that
becomes incorporated into the dead zone.  Over time the dead zone
gains sufficient mass for the disk to become self-gravitating.  The
disk state then enters the GM branch.

\subsection{Accretion rates with no steady state}

There exists a range of mass accretion rates for which no steady state
solution exists. This range is indicated by the lower shaded region in
Fig.~\ref{diagram}. This gap comes about because the GM branch
terminates at an $\dot{M}$ for which $T_{\rm c}=T_{\rm crit}$, while
the higher MRI branch starts at a larger value of $\dot{M}$ that is
also at $T_{\rm c}=T_{\rm crit}$.  As we will see in
Section~\ref{num}, a disk that has an accretion rate lying in this
range will quickly transition to either of these steady state
branches. As a consequence of such transitions, the disk undergoes
outbursts.

\subsection{Global Steady State}

We are considering a disk that undergoes accretion from an external
source at a constant accrete rate.  The $\Sigma$--$\dot{M}$ diagram
applies at each disk radius.  The range of unstable accretion rates
shown as the lower shaded region in Fig.~\ref{diagram} varies with
radius.  For a disk to be globally stable against gravo-magneto
outbursts, there should be no radius at which the disk accretion
rate lies within this range. 

\subsection{Thermal-Viscous Instability}
The case of outbursts involving the thermal-viscous instability occurs
at higher disk temperatures and accretion rates than the gravo-magneto
instability.  This regime is sketched as the well-known S-curve in the
upper portion of Fig.~\ref{diagram}.  The instability occurs along the
middle portion of the S-curve (the upper dotted line within the upper
shaded region). This middle portion consists of unstable steady state
solutions that occur between other the two stable branches. The
situation with the gravo-magneto outbursts is different because there
are no intermediate unstable steady state solutions, i.e., there are
no steady state solutions in the lower shaded region. 
For a given
accretion rate, it is possible for the disk to develop the
thermal-viscous instability at small radii and the gravo-magneto
instability further out.

\section{Time Evolution}
\label{num}

\begin{figure*}
\begin{center}
\includegraphics[width=8.8cm]{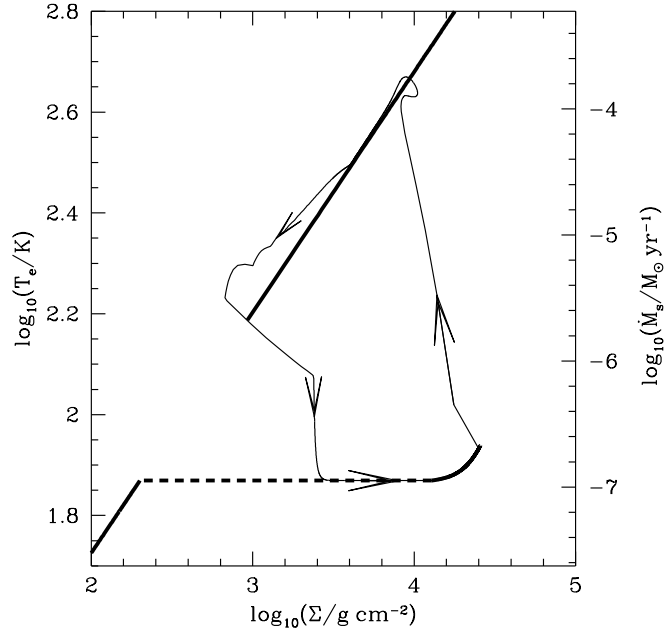}
\end{center}
\caption{The $\Sigma$--$\dot{M}_{\rm s}$ and $T_{\rm e}$ plane for a
  disk parameters listed in Section~\ref{num}. $\dot{M}_s= 3 \pi ( \nu_{\rm m} \Sigma_{\rm m} + \nu_{\rm g} \Sigma_{\rm g} )$ is the
  steady state accretion rate that applies to the steady state
  branches. The dashed line and thick lines disk solutions are defined
  in Fig.~\ref{diagram}.  The thin line shows the limit cycle of
  time-dependent disk evolution for $T_{\rm e}$ versus $\Sigma$ for an
  accretion rate on to the disk of $\dot M=10^{-6}\,\rm
  M_\odot\,yr^{-1}$ for which no steady state branch exists having
  $\dot{M} =\dot{M}_s $. }
\label{arm2}
\end{figure*}

We analyze the time evolution of a disk subject to only the
gravo-magnetic instability.  We consider a case with $R=3 \, \rm AU$,
$M=1 M_\odot$, $\dot M=10^{-6}\,\rm M_\odot\,yr^{-1}$, $Q_{\rm
  crit}=2$, $\Sigma_{\rm crit}=200\, \rm g/cm^2$, $T_{\rm crit}= 800
\,\rm K$, $\mu=2.3$, and $\alpha_{\rm m} = \alpha_{\rm g} = 0.01$.
The steady state solutions plotted in Fig.~\ref{arm2} show that no
steady state branch exists at this mass accretion rate. Therefore, we
expect outbursts in the accretion rate on to the central star.

We time-integrate the evolution equations~(\ref{sigma})
and~(\ref{temp}), together with equations (\ref{Qp})
and (\ref{Qmg-}), for a disk that extends from $R_{\rm in}=2.33\times
10^{-3}\,\rm AU$ to $R_{\rm out}=40\,\rm AU$ around a solar mass star
\citep[see e.g.][]{armitage01,martin11}. In this calculation, we
ignore the effects of the thermal-viscous instability shown by the S
curve in the upper part of Fig.~\ref{diagram}.  The grid consists of
120 points distributed uniformly in $R^\frac{1}{2}$. Material is added
to the disk at a constant rate of $\dot M=10^{-6}\,\rm M_\odot\,\rm
yr^{-1}$ at a radius of $35\,\rm AU$. At the inner radius, we impose a
zero torque boundary condition, so that there is an inward flow of gas
out of the grid and towards the central star. The flow is prevented
from leaving the outer boundary by means of a zero radial velocity
outer boundary condition.

The disk undergoes outbursts that reach a limit cycle after a few
outbursts.  In Fig.~\ref{arm2} we show the disk evolution over a limit
cycle. The cycle proceeds in a counter-clockwise sense on the
diagram.  Along the low horizontal section of the cycle, the disk is
in the dead zone state, as discussed in Section \ref{dzone}. The
surface density increases and the disk evolves to the right. The disk
then becomes self-gravitating with $Q<Q_{\rm crit}$ as a consequence
of this accumulated dead zone mass. The disk state then moves along
the gravo-magneto branch until it reaches the critical temperature for
the MRI to act, $T_{\rm crit}$. The disk then becomes
well enough ionized for MRI to act and the disk state jumps up to the
fully active MRI branch. The upward jump corresponds to the start of a
disk outburst that proceeds at a higher accretion rate. The mass then
drains out of the disk and the disk cools. As a result, the path then
moves down and to the left along the MRI branch. Once the temperature
$T_{\rm c}$ falls  below the critical value $T_{\rm crit}$, the path
moves downward. The disk once again forms a dead zone and the process
repeats itself.

\section{Discussion and Conclusions} \label{discussion}

We have explained the gravo-magneto outbursts triggered in a layered
accretion disk by means of a $\Sigma$--$\dot{M}$ diagram, along the
lines of the S-curve explanation for the thermal-viscous instability
in dwarf nova outbursts. We determined steady state disk solutions
that compose branches in the diagram.  At a given radius, there are
two types of steady-state solutions (see Fig.~\ref{diagram}). One
type, labelled MRI, involves a disk that is magnetically unstable and
turbulent at all heights. This type describes two branches of
solutions, one at low and the other at high accretion rates. Another
type, labelled GM, is gravitationally unstable and turbulent near the
mid-plane and is magnetically unstable and turbulent only near the
disk surface. This type describes one branch of solutions.  At a given
radius, there may be a range of disk accretion rates for which there
are no steady state solutions, the lower shaded region in
Fig.~\ref{diagram}.  This range of accretion rates varies with radius.
If there is a radial zone within a disk that has an accretion rate
lying in this range, then the disk will undergo outbursts in that
zone.  The outbursts are understood in terms of a limit cycle in the
$\dot{M}(\Sigma)$ diagram as shown in Fig.~\ref{arm2}.  The cycle
tracks along steady state branches of the disk solutions and along the
'dead zone' branch. The cycle jumps between states of low and high
accretion rates at the initiation and termination of an outburst. The
outburst model may be applied to accretion disks of all size scales
from circumplanetary to protoplanetary to AGN disks.
 
We have made several simplifications in our model that will be
investigated further in future work. We have ignored the
thermal-viscous instability that produces the dwarf nova outbursts
that is sketched in the upper portion of Fig.~\ref{diagram}. We
adopted a simplified opacity that does not produce this curve in the
numerical models. The thermal-viscous instability could be triggered
at smaller radii than the GM instability. The interaction between the
two instabilities may produce some interesting behavior applicable to
T Tauri stars.  The role of the thermal-viscous instability and its
S-curve has been considered by \cite{bell94} and \cite{zhu09}.

Following previous work (Armitage et al. 2001), we have taken a rather
small value for the viscosity in the active regions of $\alpha_{\rm
  m}=0.01$ and a large value for the depth of the surface magnetic
turbulence of $\Sigma_{\rm crit}= 200\, \rm g/cm^2$.  Some general
considerations suggest that $\alpha \sim 0.1$ \citep{king07}.  We note
that if the $\alpha_{\rm m}$ were higher, then the disk would be more
stable against the gravo-magneto outbursts. Similar results to those
plotted in Fig.~\ref{arm2} are obtained for $\alpha_{\rm m}=0.1$, but
with $\Sigma_{\rm crit}= 20\, \rm g/cm^2$.


We have assumed that the turbulent viscosity associated with
self-gravity, $\alpha_{\rm g}$, depends on the Toomre $Q$ parameter.
However, it has been shown in several papers that $\alpha_g$ actually
depends on the cooling rate \citep{gammie01,cossins09}.  The final
disk $Q$ value is found to be self-regulated at a value $\sim 2$.
 
We have neglected external heating of the disk and assumed that the
accretional heating dominates. For the parameters associated with the
GM instability in Fig 2, this assumption may be valid. However, more
generally external heating should be considered.

The use of a fixed value for the maximum depth of nonthermal ionization of the outer magnetic layers,
$\Sigma_{\rm crit}$, is an approximation.  A more accurate approach to
apply a critical magnetic Reynolds number required for the MRI to
operate, along the lines of \cite{matsumura03}.  In addition, we have
taken a single critical temperature for thermal ionization $T_{\rm crit}$ at the disk
mid-plane temperature above which the disk becomes MRI turbulent at
all heights. 
Instead, there may be a thin turbulent layer near the disk mid-plane that
develops at this mid-plane temperature that increases
in thickness at higher temperatures \citep{zhu09b}.

Several authors have suggested that there is a small but non-zero
turbulent viscosity in the dead zone that develops as a response to
turbulence driven in the magnetic surface layers
\citep{fleming03,turner08}.  The range of steady state disk flow
solutions is then increased \citep[e.g.,][]{terquem08}.  To some
extent, this effect will stabilize the outbursts as material can flow
through the (nearly) dead zone. However, there may be issues with the
model if the rate of flow through the dead zone approaches that
through the magnetic surface layer.  Such topics can be explored
within the framework we have described.

\section*{Acknowledgements}
We thank Jim Pringle for much advice and encouragement.  RGM thanks
the Space Telescope Science Institute for a Giacconi Fellowship.  SHL
acknowledges support from NASA grant NNX07AI72G. We thank the
referee for comments.


\label{lastpage}
\end{document}